\documentclass{sig-alternate}

\usepackage{xcolor}
\begin{document}
%

\title{On End-to-End Program Generation from User Intention\\ by Deep Neural Networks}

%
%
%
%
%

\numberofauthors{1} 
%
\author{
%
%
Lili Mou, Rui Men, Ge Li, Lu Zhang, Zhi Jin\\
       \affaddr{Software Institute, School of EECS, Peking University, Beijing 100871, P. R. China}\\
       \email{\{doublepower.mou,menruimr\}@gmail.com, \{lige,zhanglu,zhijin\}@sei.pku.edu.cn}
}

\maketitle
\begin{abstract}
This paper envisions an end-to-end program generation scenario using recurrent
neural networks (RNNs): Users can express their intention in natural language; 
an RNN then automatically generates corresponding code in a character-by-character
fashion.
 We demonstrate its feasibility through a case study and 
empirical analysis. To fully make such technique useful in practice,
we also point out several cross-disciplinary challenges, including modeling user intention,
providing datasets, improving model architectures, etc.
Although much long-term research shall be addressed in this new field, 
we believe end-to-end program generation would become
a reality in future decades, and we are looking forward to its practice.

\end{abstract}




\section{Introduction}
Imagine a following scenario in software engineering: 
There exists abundant high-quality source code, well commented
and documented, in large software repositories. A very powerful
machine (e.g., deep neural network) learns the mapping from natural
language of problem descriptions to source code. During development,
users express their intention by natural language (similar to some in the repository); 
the learning machine automatically output the desired code as the solution.

A more compelling feature is that the above process works in an
``end-to-end'' manner, which requires little, if any,
human knowledge, and is completely language independent---the only
thing needed is to represent sentences and programs as characters.
The learning machine automatically reads a natural language sentence
character-by-character to capture user intention, and then 
generates code in a similar fashion.
As learning machines differ from code retrieval systems, 
the generated code is different from any existing code, being more flexible
but maybe also vulnerable.
However, the code should be (almost) correct: It satisfies the syntax,
and implements the desired functionality. The code is usable with
a little post-editing.

Such scenario of automatic program generation has long been the dream of software engineering (SE), and
is closely related to a variety of SE tasks, e.g., algorithm discovery, 
programming assistance~\cite{review}. 
However, traditional approaches are typically weak in terms of automation and abstraction.
For example, Manna et al.~propose deductive approaches~\cite{deductive}, 
Flener et al.~inductive approaches~\cite{inductive}; these methods require human-designed specifications. 
Program generation by genetic programming~\cite{genetic1,genetic2} can automatically search the space of candidate programs (inefficiently), 
but carefully chosen mutation or crossover operations should also be provided.
Natural language programming, emerged in the past decade,
is much like ``pseudo-compiling,'' where the natural language is of
low-level abstraction \cite{natural1,natural2}.

Nowadays, software artifacts, including code and documentation, 
have become ``big data'' (e.g., Github, SourceForge).
Provided sufficient training data of code with
corresponding comments and documents, it is
possible in principle to train a generative model
of programs based on natural language.
At the meantime, the natural language processing (NLP)
community is witnessing significant breakthroughs 
and amazing results in various tasks
 including question answering~\cite{QA},
machine translation~\cite{seq2seq},
or image-caption generation~\cite{caption}. 
These cross-disciplinary advances bring new opportunities
for automatic program generation.

In this paper, we investigate, by a case study in Section~\ref{case}, 
the feasibility of generating executable, functionally coherent 
code by recurrent neural networks (RNNs);
empirical analysis reveals the mechanism how RNNs could accomplish the goal.
We also envision several scenarios where such techniques may help real-world 
SE tasks, and address long-term research challenges (Section~\ref{sEnvision}). 
Although we concede there still remains a long way 
before end-to-end program generation can be used in SE practice, we believe
it would become a reality in future decades.

\section{A Case Study}\label{case}

\subsection{The Model of Recurrent Networks}

Among a variety of machine learning methods, the deep neural network (also known as \textit{deep learning})
is among recent groundbreaking advances, featured by its ability of learning
highly complicated features automatically \cite{DL}.

\begin{figure}[!t]
\centering
\includegraphics[width=.45\textwidth]{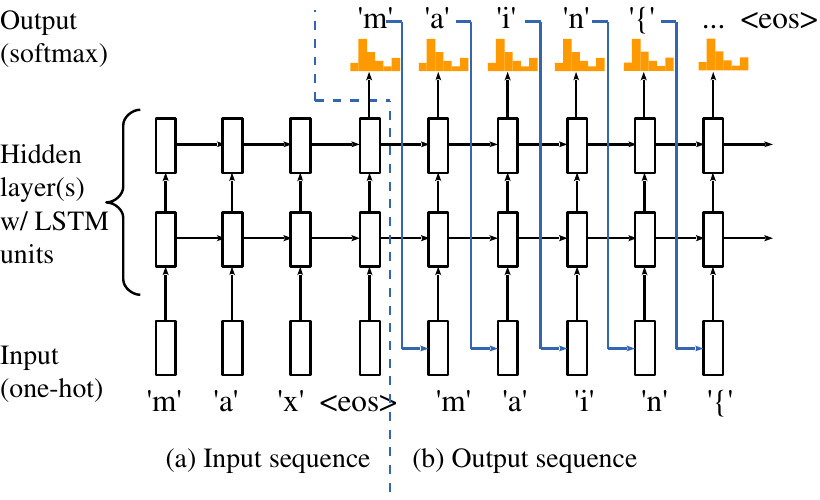}
\vspace{-.5cm}
\caption{A sequence to sequence recurrent neural network, adapted from \cite{seq2seq}.
(a) Input sequence; (b) Output sequence.}\label{fRNN}
\end{figure}

For end-to-end program generation, we prefer the recurrent neural networks (RNNs), which 
are suitable for modeling time-series data 
(e.g., a sequence of characters) by its iterative nature. An RNN typically keeps one or a few hidden layers, 
changing over each (discrete) time step according to input data. This process is delineated in Figure~\ref{fRNN}.

Theoretical analysis shows that recurrent neural networks are equivalent to Turing machines~\cite{turing}. However, training RNNs in early years was difficult because of the \textit{gradient blowup or vanishing} problem \cite{difficulty}. Long short term memory (LSTM) units~\cite{lstm}, or gated units \cite{gru} are designed to balance between retaining the previous state and memorizing new information at the current time step, making RNNs much easier to train.

On this basis, Sutskever et al.~design an RNN model for sequence to sequence generation \cite{seq2seq}. 
The idea is to first read an input sequence, ended with a special symbol, \verb|<eos>| (end of sequence), 
depicted by Figure~\ref{fRNN}a. For output, the RNN applies a \verb|softmax| layer at each time step, predicting the probability that each symbol\footnote{A symbol may refer to a word or a character according to the granularity in a certain application.} may occur at the current step; the symbol with the highest probability is chosen, and fed to the network as input at the next time step. This process is done iteratively until the special symbol, \verb|<eos>|, is generated by the network (Figure~\ref{fRNN}b).

Such RNN architecture can be applied to sequences of different granularities, e.g., word-level, sub-word level, etc. 
In particular, character-level RNN generative models have unexpectedly achieved remarkable and somewhat amazing performance. Successful applications include generating texts, music, or even Linux-like \verb|C| code~\cite{unreasonable}. Empirical studies show that RNNs are particularly good at modeling syntax aspects, e.g., parenthesis pairing, indentation, 
etc~\cite{visualize}. It works much like a push-down automata, but seems less capable of capturing semantics---the Linux-like code generated, for example, is plausible, but cannot be compiled and lacks coherence in functionality.

We are therefore curious whether RNNs can generate executable, functionally coherent source code, which is an essence to benefit real-world software engineering tasks. 

To accomplish this goal, we leverage a dataset from a pedagogical programming online judge (OJ) system,\footnote{http://programming.grids.cn} intended for the undergraduate course,
\textit{Introduction to Computing}.
The OJ system comprises different programming problems.
Students submit their source code to a specific problem, and the OJ system
judges its validity automatically (via running). 
We notice that programs corresponding to a specific problem have
\textit{exactly the same} functionality, which makes the dataset 
particularly suitable, as a first trial, for training neural networks to generate functionally
coherent programs.

\begin{figure*}[!t]
\hspace*{0cm}\includegraphics[width=\textwidth]{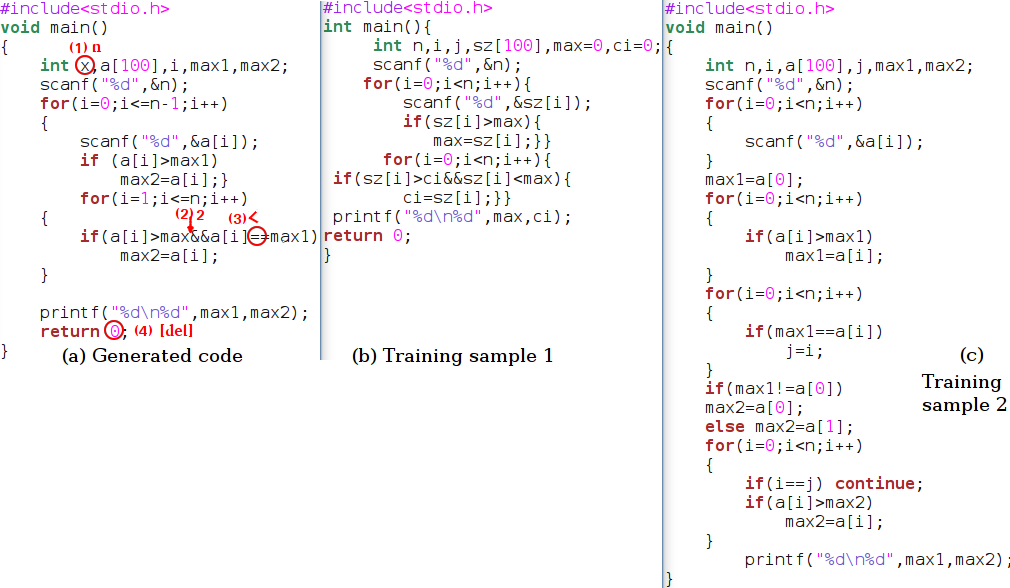}
\begin{minipage}{.63\textwidth}
\vspace*{-4.5cm}
\caption{(a) Code generated by RNN. The code is almost correct
except 4 wrong characters (among $\sim$280 characters in total), highlighted in the figure. 
(b) Code with the most similar structure in the training set, detected by
{\tt ccfinder}. (c) Code with the most similar identifiers in the training set,
also detected by {\tt ccfinder}. Note that we preserve all indents, spaces and line feeds.
The 4 errors are (1) The identifier ``{\tt x}'' should be ``{\tt n}''; (2) ``{\tt max}'' should be 
``{\tt max2}''; (3) ``{\tt ==}'' should be ``{\tt <}''; (4) return type should be void.
}\label{fResult}
\end{minipage}
\vspace*{-.5cm}
\end{figure*}

We fed the network with 4 different programing problems,
each containing more than 500 different source code samples.
After preprocessing, a program was preceded by a brief comment, e.g., ``find the maximum and second maximum numbers,'' serving as the input sequence (Figure~\ref{fRNN}a).\footnote{
Entire dataset and configurations are available on our website.
http://sites.google.com/site/rnngenprogram} What follows is a program that solves the particular problem, serving as the output sequence (Figure~\ref{fRNN}b).
Figures~\ref{fResult}b and \ref{fResult}c further illustrate two training samples of the aforementioned programming problem.
 
\subsection{Result and Analysis}

Figure~\ref{fResult}a is a sample code generated by RNN.
Through a quick analysis, we find that the code is almost executable: with a little post-correction
of 4 characters among $\sim$280, the program is compilable and functionally correct.

We would answer a very fundamental question: Is RNN generating code by
simply memorizing a particular training sample? If this were the case,
RNN would just work in a copy-and-paste fashion, which degrades the problem
to a trivial case. 

By examining the training data, we observe that there does NOT exist a same program
in the training set, which rules out the possibility that RNN works by exact memorizing. We further use {\tt ccfinder}\footnote{http://www.ccfinder.net/} to detect most similar
code in the training set. Two are shown in Figure~\ref{fResult}, and the results are particularly interesting.
We provide our explanation regarding several aspects of a program as follows.
\begin{itemize}
\item \textbf{Structure.} Figure~\ref{fResult}b shows the most
similar code in structure. The generated code implements the same algorithm---scanning the array twice to 
find the maximum and second maximum
numbers respectively. Notice, however, the two structures (abstract syntax trees, say) are not exactly the same as there are differences in 
variable definitions. A more interesting
detail is that RNN has recognized ``{\tt i<n}'' and ``{\tt i<=n-1}'' are equivalent
in the {\tt for} loop, and
that it does not follow exactly the sample code (b) in the training set but remains correct.
\item \textbf{Variable IDs.} The training sample with the most similar variable IDs is shown in Figure~\ref{fResult}c. Our generated code uses the same ID,
{\tt a}, for the array, and {\tt max1}, {\tt max2} to cache the two wanted numbers;
but later, the structure diverges. Nevertheless, our network is aware of the variable IDs
it has generated, and remains coherent until the very end of the program.

\item \textbf{Style.} We find no particular training samples having the same code style in terms of indents, line feeds, etc. It makes much sense because
the training programs are written by junior programmers, who may not follow
standard style convention, and thus the network has no idea about the ``right''
style. However, as all training samples are ``correct'' programs, our network
has little difficulty in learning the syntax of {\tt C} programs as the generated code
can almost be compiled.
\end{itemize}

Through the above analysis, we gain a basic idea on how RNN is able to generate programs.
The RNN first recognizes the brief comment, ``find the maximum and second maximum numbers,'' which precedes the code as input. We would like to point out that, in this experiment, the RNN does not understand the meaning of this sentence; but via reading the brief comment, the RNN 
switches its hidden states to generate code of the functionality in need.
For each functionality, RNN is aware of different aspects of a possible program, including  structures, IDs, etc. When generating, it chooses the most likely character conditioned on the previous characters, also conditioned on the input. In particular,
the RNN does have the ability to mix different structures and IDs but remain (almost) coherent.

\section{Prospectives \& Road Map}\label{sEnvision}

While simple and preliminary, 
our case study and analysis provide illuminating pictures on
end-to-end program generation with deep neural networks. 
We point out several scenarios where deep learning
can benefit real-world SE practice, which are also 
research topics in long-term studies.

\begin{itemize}
\item Understanding changeable user intention. 
The current case study shows RNN's ability of recognizing 
certain intention and generating corresponding code. 
In SE practice, however, we are oftentimes facing changeable requirement from users.
To address the problem, a direct extension is to train a parametric code generator with arguments
(e.g., file names, protocols) implicitly expressed using natural language.
To tackle a more challenging prospective, 
we might first train a network to generate different 
``primitive'' code snippets, and then ``glue'' them together.
For instance, if a network has learned to write code of 
finding the maximum number, and also
of finding the minimum number, then it shall be possible to generate these two snippets
subsequently if it reads an instruction ``find the maximum and minimum numbers.''

\item Incorporating multiple sources of user intention. 
When developing software, programmers usually find their code
dependent to context (e.g., previously defined variables, existing API call sequences) in addition to
the functionality in need. 
In such scenarios, we might train a network
to fill missing blocks of code. While we admit that code completion in general
could hardly make any sense, we think this problem is mostly realistic in some
task-specific scenarios.
For example, a typical way of reading a \verb|txt| file in \verb|Java| involves
creating \verb|FileReader|, \verb|BufferedReader|, reading lines in the file,
closing the file, and also catching exceptions. Such standard pipelines might
be generated automatically by neural networks, provided context code.

\end{itemize}

Despite the promising future of using RNNs to generate source code, 
efforts shall be made from multiple disciplines including SE, NLP and machine 
learning communities. Most important questions in the SE community
are defining user's intention and providing datasets for training. 
How can we specify the functionality that we want to generate? 
How can we specify the arguments of a function? 
How can we collect the dataset which
is not only large and informative enough for training, but also clean enough for not including too
much noise? These are among the open questions.
The NLP and machine learning communities, on the other hand, are continuously improving
neural architectures. Attention-based networks \cite{attention,caption}, for example, 
are proposed recently to mitigate the problem
of long input sequences that cannot be composed into a fixed-size vector.
More studies are still needed in terms of understanding the memory capacity of RNNs, generating data with more coherent semantics, or even revising generated data, etc.

We concede that using RNNs to generate programs differs significantly
from writing programs by humans. It appears unrealistic currently to train any learning machine, including deep neural networks, to fully understand either natural languages or programming languages. 
However, supported by existing evidence in the literature and the case study in this paper,
we deem end-to-end program generation shall be possible in the future.

\section{Related Work in Deep Learning For Program Analysis}

Recent years have witnessed the birth of program analysis based on deep neural networks.
Our previous work learns programs' vector representations, serving as a pretraining phrase in deep learning~\cite{building}; 
we also propose tree-based convolutional neural networks to classify programs by functionality and detect source code of certain patterns~\cite{tbcnn}. 
Zaremba et al.~use RNNs to estimate the output of a restricted python program \cite{execute}. 
 Allamanis et al.~leverage vector representations to suggest method names \cite{suggest}.
All the above models are \textit{discriminative}, by which we mean the tasks
can be viewed as a classification problem. 
Karpathy et al.~train an RNN-based language model on \verb|C| code, which maximizes the joint probability of a program~\cite{visualize}. Different from the above studies, 
this paper investigates whether neural models can synthesize executable, functionally coherent programs, which demands more need in matching users' intention and capturing internal structures of source code.

\section{Conclusive Remarks}
In this paper, we trained a recurrent neural network (RNN) 
to generate (almost) executable, functionally coherent source code.
Our initial work has demonstrated the possibility 
of automatic end-to-end program generation.
Through analyzing the RNN's mechanism, we envisioned
several scenarios where such techniques can be applied in 
software engineering tasks in future decades.
We call for studies from multiple disciplines to further
address this new research direction.



\bibliographystyle{abbrv}

\bibliography{sigproc}

\begin{thebibliography}{10}

\bibitem{suggest}
M.~Allamanis, E.~Barr, C.~Bird, and C.~Sutton.
\newblock Suggesting accurate method and class names.
\newblock In {\em ESEC/FSE}, 2015.

\bibitem{gru}
K.~Cho, B.~van Merri{\"e}nboer, D.~Bahdanau, and Y.~Bengio.
\newblock On the properties of neural machine translation: Encoder-decoder
  approaches.
\newblock {\em arXiv preprint}, 1409.1259, 2014.

\bibitem{attention}
J.~Chorowski, D.~Bahdanau, D.~Serdyuk, K.~Cho, and Y.~Bengio.
\newblock Attention-based models for speech recognition.
\newblock {\em arXiv preprint}, 1506.07503, 2015.

\bibitem{natural2}
A.~Cozzie and S.~T. King.
\newblock Macho: {W}riting programs with natural language and examples.
\newblock {\em Technical report, University of Illinois at Urbana-Champaign},
  2012.

\bibitem{inductive}
P.~Flener and D.~Partridge.
\newblock Inductive programming.
\newblock {\em Automated Softw. Engineering}, 8(2):131--137, 2001.

\bibitem{review}
S.~Gulwani.
\newblock Dimensions in program synthesis.
\newblock In {\em Proc. ACM SIGPLAN Symposium on Principles and Practice of
  Declarative Programming}, 2010.

\bibitem{genetic2}
T.~Helmuth and L.~Spector.
\newblock General program synthesis benchmark suite.
\newblock In {\em Proc. Genetic and Evol. Comput. Conf.} ACM, 2015.

\bibitem{lstm}
S.~Hochreiter and J.~Schmidhuber.
\newblock Long short-term memory.
\newblock {\em Neural Comput.}, 9(8):1735--1780, 1997.

\bibitem{turing}
H.~Hy{\"o}tyniemi.
\newblock Turing machines are recurrent neural networks.
\newblock {\em Proc. STeP}, 1996.

\bibitem{unreasonable}
A.~Karpathy.
\newblock The unreasonable effectiveness of recurrent neural networks.
\newblock {\em http://karpathy.github.io/ 2015/05/21/rnn-effectiveness/}, 2015.

\bibitem{visualize}
A.~Karpathy, J.~Johnson, and F.~Li.
\newblock Visualizing and understanding recurrent networks.
\newblock {\em arXiv preprint}, 1506.02078, 2015.

\bibitem{natural1}
R.~Kn{\"o}ll and M.~Mezini.
\newblock Pegasus: {F}irst steps toward a naturalistic programming language.
\newblock In {\em OOPSLA}, 2006.

\bibitem{QA}
A.~Kumar, O.~Irsoy, J.~Su, et~al.
\newblock Ask me anything: {D}ynamic memory networks for natural language
  processing.
\newblock {\em arXiv preprint}, 1506.07285, 2015.

\bibitem{DL}
Y.~LeCun, Y.~Bengio, and G.~Hinton.
\newblock Deep learning.
\newblock {\em Nature}, 521(7553):436--444, 2015.

\bibitem{deductive}
Z.~Manna and R.~Waldinger.
\newblock A deductive approach to program synthesis.
\newblock {\em ACM Trans. Programming Languages and Syst.}, 2(1):90--121, 1980.

\bibitem{tbcnn}
L.~Mou, G.~Li, Z.~Jin, L.~Zhang, and T.~Wang.
\newblock {TBCNN: A} tree-based convolutional neural network for programming
  language processing.
\newblock {\em AAAI Workshop}, 2015.

\bibitem{building}
L.~Mou, G.~Li, Y.~Liu, H.~Peng, Z.~Jin, Y.~Xu, and L.~Zhang.
\newblock Building program vector representations for deep learning.
\newblock {\em arXiv preprint}, 1409.3358, 2014.

\bibitem{difficulty}
R.~Pascanu, T.~Mikolov, and Y.~Bengio.
\newblock On the difficulty of training recurrent neural networks.
\newblock {\em arXiv preprint}, 1211.5063, 2012.

\bibitem{seq2seq}
I.~Sutskever, O.~Vinyals, and Q.~Le.
\newblock Sequence to sequence learning with neural networks.
\newblock In {\em NIPS}, 2014.

\bibitem{genetic1}
W.~Weimer, T.~Nguyen, C.~Le~Goues, and S.~Forrest.
\newblock Automatically finding patches using genetic programming.
\newblock In {\em ICSE}, 2009.

\bibitem{caption}
K.~Xu et~al.
\newblock Show, attend and tell: {N}eural image caption generation with visual
  attention.
\newblock {\em arXiv preprint}.

\bibitem{execute}
W.~Zaremba and I.~Sutskever.
\newblock Learning to execute.
\newblock {\em arXiv preprint}, 1410.4615, 2014.

\end{thebibliography}

%
%
\end{document}